\documentclass[12pt]{ronbun}

\usepackage{color,amsmath,amssymb,graphicx,epsfig}
\usepackage{cite}
\usepackage{bm}
\usepackage{dcolumn}

\newcommand{\nn}{\nonumber}

\numberwithin{equation}{section}

\begin{document}

\begin{flushright}

\parbox{3.2cm}{
{KEK-TH-935 \hfill \\
{\tt hep-th/0402028}}\\
\date
 }
\end{flushright}

\vspace*{0.7cm}

\begin{center}
 \Large\bf Effective Potential in PP-Wave Geometry
\end{center}

\vspace*{1.0cm}

\centerline{\large Kunihito Uzawa$^{\dagger a}$ and   
Kentaroh Yoshida$^{\ast b }$}

\begin{center}
$^{\dagger}$\emph{Graduate School of Human and Environmental Studies,
\\ Kyoto University, Kyoto 606-8501, Japan.} \\
$^{\ast}$\emph{Theory Devision, 
 High Energy Accelerator Research Organization (KEK),
\\ Tsukuba, Ibaraki 305-0801, Japan.} \\
{\tt E-mail:~$^{a}\,$uzawa@phys.h.kyoto-u.ac.jp \\
\quad $^{b}\,$kyoshida@post.kek.jp}\\

\vspace{0.2cm}
\end{center}

\vspace*{1.0cm}

\centerline{\bf Abstract}

We calculate effective potentials in scalar field theories 
on the maximally supersymmetric pp-wave background in ten dimensions. 
For this purpose we have to work in the light-cone formulation, and
hence we introduce two methods to compute them in the light-cone frame. 
One is to use the Yan's formula for evaluating one-loop correction terms. 
The other is to introduce a cut-off for the light-cone momentum. 
These methods are also confirmed in the case of Minkowski spacetime.

\vspace*{1.2cm}

\thispagestyle{empty}
\setcounter{page}{0}
\newpage 

\section{Introduction}

In the past years, pp-wave backgrounds are very focused upon
 \cite{Kowalski-Glikman,Blau1}.  String theories on pp-waves are exactly
 solvable \cite{M,MT} and play an important role in studies of
 AdS/CFT correspondence \cite{BMN}.  It is also an interesting problem
 to apply plane-wave geometries for cosmological models. In the recent
 progress, Penrose limits were applied for a variety of the space-time
 geometries, such as ${\rm AdS}_{\rm p} \times {\rm S}^{\rm q}$
 \cite{Blau2,Blau3}, AdS black
 holes \cite{Marolf1,Hubeny,Brecher:2002bw,Marolf:2002bx,Li}, and
 various brane configurations (for example,
 \cite{Blau2,Blau3,Oz:2002ku,Fuji:2002vs,Ito:2003vu}).  More recently,
 the plane-wave type Friedmann-Robertson-Walker (FRW) solution was
 proposed in \cite{Papadopoulos:2002bg}.  This model is not static while
 it is exactly solvable as a feature of plane-wave backgrounds.  Another
 type of time-dependent plane-wave background was also found \cite{BO}
 and the string theory on this background was studied \cite{BOPT}. In
 addition, an example of this type of background in eleven dimensions
 with extra supersymmetries was proposed in \cite{BMO} and the M-theory
 on this background was studied in \cite{SaYo}.  Motivated by these
 recent attempts, we are interested in constructing other cosmological
 models connected to plane-wave geometry. When we intend to construct
 such models, it would be a first step to consider scalar field theories
 on pp-wave backgrounds.  So far the scalar field theories on pp-waves
 are discussed in \cite{Marolf1,Brecher:2002bw,Marolf:2002bx,Berenstein}
 and scalar propagator was calculated \cite{Mathur}.  However, quantum
 aspects of field theories on pp-waves have not been revealed yet well.

In this paper, we will calculate effective potentials in scalar field
theories on the ten-dimensional maximally supersymmetric pp-wave
background \cite{Blau1}. We need the methods to calculate effective
potentials in the light-front formulation.  However, such methods have
some difficulties and so have been discussed still now (for example, see
\cite{Heinzl}).  Thus, we will discuss two methods to calculate
effective potentials in the light-front formulation as a setup.  One is
to use the Yan's formula \cite{Yan}, and the other is to introduce the
cut-off for the light-cone momentum.  By using these methods, we will
calculate effective potentials in scalar field theories in
four-dimensional Minkowski spacetime in order to confirm that our
methods lead to the well-known results derived by the dimensional
regularization and minimal subtraction scheme.  These methods can be
used to calculate the effective potential in the case of pp-wave
backgrounds.  The resulting effective potential is similar to that in
the two-dimensional Minkowski spacetime up to the numerical coefficient.

This paper is organized as follows: In section 2, we present 
two methods to calculate effective potentials in the 
light-cone frame. By using these methods 
 effective potentials in scalar field 
theories in  four-dimensional flat space in order to confirm our methods. 
In section 3, we calculate effective potentials in 
scalar field theories in the ten-dimensional maximally
 supersymmetric pp-wave background. 
Section 4 is devoted to a conclusion and discussions. 
In appendix, we will summarize several useful formulae for the calculations in this paper.   

\section{Calculi of Effective Potentials in Light-Front Formulation}
\label{sec;min}

In this section we will discuss calculi of effective potentials 
in the light-cone formulation in the case of four-dimensional Minkowski 
spacetime, though our purpose in this paper is to compute them in the
pp-wave case. It is inevitable in the pp-wave case 
to work in the light-cone formulation, but 
the computation of effective potentials in the light-cone frame contains
some technical subtleties. Hence, it would be surely available to
introduce carefully the computation method. 

To begin with, we will briefly review the zeta-function regularization,
which is a standard method to calculate effective potentials. 
Then we will introduce two methods to calculate effective potentials in
the light-cone formulation. 

\subsection{Standard Calculus of Effective Potentials}

Let us consider the effective potential in a 
$d$-dimensional scalar field theory, by using the zeta-function
regularization method. 
The action and partition function are, respectively, given by
\begin{equation}
I[\phi] = \int\! d^d x\, \left\{ - \frac{1}{2}g^{\mu\nu} 
\partial_{\mu}\phi\;\partial_{\nu}\phi - V(\phi) \right\}\,,
\hspace{1cm}
Z=\int {\cal D}\phi\;{\rm e}^{iI[\phi]}\,,
\end{equation}
where $V(\phi)$ is a given potential. 
When we expand the field variable $\phi$ around the constant 
classical background
$\phi_{\rm c}$ as $\phi = \phi_{\rm c} + \delta\phi$, the effective
 action 
$\Gamma[\phi_{\rm c}]$ is 
represented by 
\begin{eqnarray}
i\Gamma[\phi_{\rm c}] \equiv \ln Z &=& i\int\! d^dx\,V(\phi_{\rm c}) 
   +\ln\left\{\int{\cal D}\delta\phi\;
     \exp{\left(iI_{\rm q}[\delta\phi]\right)}\right\}\,,
          \label{eq;3-2-2} \nonumber \\
I_{\rm q}[\delta\phi] &=& - \frac{1}{2}\int\! d^d x\, 
        \delta\phi\left(\Box_{(d)} + V''_{\rm c} \right)\delta\phi\,,
\end{eqnarray}
where we defined  
$\Box_{(d)} \equiv -\,g^{\mu\nu}\partial_{\mu}\,\partial_{\nu}$ and 
$V''_{\rm c} \equiv d^2 V/d\phi^2|_{\phi=\phi_{\rm c}}$. 
By performing the path integration, 
the effective potential is obtained as 
\begin{eqnarray}
\Gamma[\phi_{\rm c}] &\equiv& - 
\int\! d^dx\, V_{\rm eff}[\phi_{\rm c}]=-V_{d}\cdot
  V_{\rm eff}[\phi_{\rm c}]
   \,,\nn\\
V_{\rm eff}(\phi_{\rm c}) &=& V(\phi_{\rm c})-\frac{i}{2V_{d}}\,
        {\rm Tr}\ln\Big\{M^{-2}
        \left(\Box_{(d)}+V''_{\rm c}\right)\Big\}\,.   
    \label{eq;eff2}
\end{eqnarray}
Here we introduced a positive scale-parameter $M$ with a mass dimension
to make an argument dimensionless.  We can evaluate the trace-log term
by introducing the generalized zeta-function:
\begin{equation}
\zeta_{\rm Min}(s) = \int\!\frac{d^dk}{(2\pi)^d}\, 
         \left(-k^2+V''_{\rm c}\right)^{-s}\,,
         \label{eq;zetam4sum}
\end{equation}
and then the effective potential is expressed as 
\begin{equation}
V_{\rm eff}\left(\phi_{\rm c}\right)=V\left(\phi_{\rm c}\right)
            +\frac{i}{2}
            \left(\frac{d\:\zeta_{\rm Min}(s)}{ds}\Biggr|_{s=0}
            +\zeta_{\rm Min}(0)\cdot\ln M^2\right)\,.
            \label{eq;zetaeff}
\end{equation}
The evaluation of the generalized zeta-function in the usual formulation
is an easy task.  As an example, the effective potential in
four-dimensional Minkowski spacetime is given by
\begin{eqnarray}
V_{\rm eff}\left(\phi_{\rm c}\right) 
          =V\left(\phi_{\rm c}\right)
          +\frac{1}{4(4\pi)^{2}}
          \left(V''_{\rm c}\right)^2
          \left\{\ln\left(\frac{V''_{\rm c}}
          {M^2}\right)
           -\frac{3}{2}\right\}\,. 
       \label{eq;defineEP}
\end{eqnarray}

But, the situation is quite different in the light-cone formulation,
because the evaluation of the generalized zeta-function
(\ref{eq;zetam4sum}) is different from that in flat case, and it is
slightly difficult and technical.  
In the following subsections we will
present new methods for calculating the one-loop correction term in the
light-cone frame, which are based on the zeta-function regularization
method. Namely, we present the method to evaluate the generalized
zeta-function in the light-cone formulation. 
One is to use the Yan's formula\cite{Yan}, and
another is to introduce the cut-off for the light-cone momentum. 
These methods will be applied to the pp-wave case.

\subsection{Calculus with Yan's Formula}

We now present a method to evaluate the generalized zeta-function 
in the light-cone frame. One of the subtleties is 
a Wick rotation in the light-cone formulation. 
However we can avoid a Wick rotation by using the Yan's formula. 
This method is a slight generalization 
of the recent impressive work of Heinzl \cite{Heinzl}. 

Let us consider the case of four-dimensional Minkowski spacetime 
with the light-cone coordinates $x^{\pm} = (x^0 \pm x^3)/\sqrt{2}$.  
Now the generalized zeta function is written as 
\begin{equation}
\zeta_{\rm Y4}(s)=\int^{\infty}_{-\infty}\frac{dk_{+}}{2\pi}
          \int^{\infty}_{-\infty}\frac{dk_{-}}{2\pi}
          \int^{\infty}_{-\infty}\frac{d^2k}{(2\pi)^2}\,
          \left(-2k_{+}\:k_{-} + k^2+V_{\rm c}''\right)^{-s}\,. 
           \label{eq;zetayan}
\end{equation}
The integration for the transverse momenta leads to 
\begin{equation}
\zeta_{\rm Y4}(s)=\frac{\pi}{(2\pi)^{4}}\frac{\Gamma(s-1)}{\Gamma(s)}
          \int^{\infty}_{-\infty}dk_{-}\int^{\infty}_{-\infty}dk_{+}\,
          (-2)^{-s+1}\left(k_{+}\:k_{-}-\frac{V_{\rm c}''}
          {2}\right)^{-s+1} \,.
\end{equation}
The integral for $k_+$ needs a careful treatment of  
single pole $k_+=V_{\rm c}''/2k_-$\,. We shift the pole by $\epsilon$ 
and use the Yan's formula\,,
\begin{eqnarray}
\int^{\infty}_{-\infty}\!\!dk_{+}\,
     \left(k_{+}\:k_{-}-m^2+i\epsilon\right)^{-s} &=& 2\pi i\,\delta(k_-)
          \frac{(-1)^{s}}{s-1}\left(m^2\right)^{-s+1}\,,
          \label{eq;yan}
\end{eqnarray}
and then 
Eq.\,(\ref{eq;zetayan}) is rewritten as  
\begin{eqnarray}
\zeta_{\rm Y4}(s)&=& i\frac{1}{(4\pi)^{2}}\frac{\Gamma(s-2)}{\Gamma(s)}
          \left(V_{\rm c}''\right)^{-s+2}\,.
        \label{eq;finalyan}
\end{eqnarray}
Note that the pole of the integrand in the l.h.s of Eq.\,(\ref{eq;yan})
is at infinity when $k_-=0$, and this fact leads to $\delta(k_-)$\,.

As a matter of course, 
the final expression (\ref{eq;finalyan}) calculated in 
the light-cone frame leads to 
the standard result. It should be noted that 
a Wick rotation is not needed as noted in \cite{Yan}.  
In the light-cone formulation the notion of Wick rotation 
is subtle since the light-cone time plays a main role 
instead of the usual time-coordinate. But 
such a subtlety can be avoided by using the Yan's formula. 

We can apply the method introduced here to the pp-wave case. 
In the next subsection we will discuss another method to calculate  
effective potentials in the light-cone frame.

\subsection{Calculus with the Light-Cone Momentum Cut-Off}

We shall present another calculus of effective potentials in the
light-cone frame. Here we do not use the Yan's formula but introduce
a cut-off for the light-cone momentum. 

To begin with, we rewrite the one-loop correction term as    
\begin{eqnarray}
{\rm Tr}\ln\left\{\Box_{\rm (LC)} 
     + V''_{\rm c}
          \right\}
    &=& \int\!\!\frac{dk_+\,dk_-}{(2\pi)^2}\int\!\!\frac{d^2k}{(2\pi)^2}\,
       \ln\left(-2k_+ k_- + k^2 + V_{\rm c}''\right)\,.
     \label{eq;loopyan}
\end{eqnarray}
Then, by following the work of Heinzl\cite{Heinzl}, 
let us introduce a cut-off for the light-cone momentum $k_-$ as 
\begin{eqnarray}
V''_{\rm c}/\Lambda \leq |k_-| \leq \Lambda\,,
\end{eqnarray}
while the standard cut-off is given by $m^2/\Lambda\leq |k_-|\leq
\Lambda$\,. The one-loop correction term (\ref{eq;loopyan}) can be
rewritten as
\begin{eqnarray}
{\rm Tr}\ln\left\{\Box_{\rm (LC)} 
     + V''_{\rm c}
          \right\} 
    &=& \int^{\infty}_{-\infty}\!\!\,\frac{dk_+}{2\pi}
        \int^{\infty}_{-\infty}\!\!\,\frac{d^2k}{(2\pi)^2}\,
     \left\{\int^{\Lambda}_{V''_{\rm c}/\Lambda}\!\!\,\frac{dk_-}{2\pi}
       \ln\left(-2k_+ k_- + k^2 + V_{\rm c}''\right)\right.\nonumber\\
    & &\left. \quad  +\int^{-V''_{\rm c}/\Lambda}_{-\Lambda}\!\!\,
      \frac{dk_-}{2\pi}
       \ln\left(-2k_+ k_- + k^2 + V_{\rm c}''\right)\right\}  \\
    &=& \int^{\infty}_{-\infty}\!\!\,\frac{dk_+}{2\pi}
        \int^{\Lambda}_{V''_{\rm c}/\Lambda}\!\!\frac{dk_-}{2\pi}
        \int^{\infty}_{-\infty}\!\!\,\frac{d^2k}{(2\pi)^2}\,
       \ln\left\{-\left(2k_+ k_-\right)^2 
        + \left(k^2 + V_{\rm c}''\right)^2\right\}\,. \nn
\end{eqnarray}
For this expression of one-loop correction, 
the generalized zeta-function is 
\begin{eqnarray}
\zeta_{\rm LC4}(s)&=&\int^{\infty}_{-\infty}\frac{dk_{+}}{2\pi}
          \int^{\Lambda}_{V_{\rm c}''/\Lambda}\frac{dk_{-}}{2\pi}
          \int^{\infty}_{-\infty}\frac{d^2k}{(2\pi)^2}
          \left\{-\left(2k_{+}\:k_{-}\right)^2
          +\left(k^2+V_{\rm c}''\right)^2\right\}^{-s}\,.
\end{eqnarray}
By performing the analytic continuation $k_+\rightarrow ik_+$\footnote{
The ambiguity of the sign exists, but we choose the sign which 
leads to the same result as in the standard calculation.}, 
we can carry out the integration for $k_+$\,. 
As the result, we obtain the following expression 
\begin{eqnarray}
\zeta_{\rm LC4}(s)
          &=&i\,\frac{\sqrt{\pi}\:\Gamma\left(s-\frac{1}{2}\right)}
               {(2\pi)^{4}\,\Gamma(s)}
              \int^{\infty}_{-\infty}d^2k
              \left(k^2+V_{\rm c}''\right)^{-2s+1}
              \int^{\Lambda}_{V_{\rm c}''/\Lambda}\frac{dk_{-}}{2k_-} 
             \nonumber\\
          &=&-i\,\frac{\pi^{3/2}\Gamma\left(s-\frac{1}{2}\right)}
              {2(2\pi)^{4}\,\Gamma(s)}\,
              \frac{\Gamma\left(2s-2\right)
              \left(V_{\rm c}''\right)^{-2s+2}}
              {\Gamma\left(2s-1\right)}
              \ln \left(\frac{V_{\rm c}''}{\Lambda^2}\right)\,,
\end{eqnarray}
and the derivative of it is 
\begin{eqnarray}
\zeta'_{\rm LC4}(s)
          &=&-i\,\frac{\pi^{3/2}\Gamma\left(s-\frac{1}{2}\right)\psi(s)}
              {2(2\pi)^{4}\,\Gamma(s)}\,
              \frac{\Gamma\left(2s-2\right)
              \left(V_{\rm c}''\right)^{-2s+2}}
              {\Gamma\left(2s-1\right)}
              \ln \left(\frac{V_{\rm c}''}{\Lambda^2}\right)\,.
\end{eqnarray}
Here $\psi(s)$ is a polygamma function. 
The $s\rightarrow 0$ limits of them are, respectively,  
\begin{eqnarray}
\lim_{s\rightarrow 0}\zeta_{\rm LC4}(s)
    = 0\,, \hspace{1cm} \lim_{s\rightarrow 0}\zeta_{\rm LC4}'(s)
    =-i\,\frac{1}{2(4\pi)^{2}}\,\left(V_{\rm c}''\right)^{2}\,
         \ln\left(\frac{V_{\rm c}''}{\Lambda^2}\right)\,,
\end{eqnarray}
and hence the resulting effective potential is  
\begin{eqnarray}
V_{\rm eff}\left(\phi_{\rm c}\right)
       =V\left(\phi_{\rm c}\right)
          +\frac{1}{4(4\pi)^{2}}\,
          \left(V''_{\rm c}\right)^2\,
          \ln\left(\frac{V''_{\rm c}}{\Lambda^2}
          \right)\,.
   \label{eq;LC4}
\end{eqnarray}
Now we should notice the ambiguity of multiplicative constant 
of the cut-off $\Lambda$. When we rescale $\Lambda$ as 
$\Lambda^2\rightarrow \Lambda^2\:{\rm e}^{3/2}\equiv M^2$,  
the effective potential is identical with the standard result. 

The method presented here can be also applicable to the pp-wave case. 
This kind of application will be discussed in the next section.

\section{Effective Potentials on PP-Wave Background}
\label{sec;pp}

In this section we consider the effective potential in scalar field
theories on the ten-dimensional maximally supersymmetric background 
\cite{Blau1}. 
This background is described by 
\begin{equation}
ds^2=-2dx^{+}dx^{-} - \mu^2r^2\left(dx^+\right)^2
       +\sum_{i=1}^{8} dx_idx^i\,,    
        \label{eq;pp-metric}
\end{equation}
where $r^2= \sum_{i=1}^8 (x^i)^2$ 
and the parameter $\mu$ is related to the constant flux of Ramond-Ramond 
four-form  
\begin{equation}
F_{+1234}=F_{+5678}=4\mu\,. 
\end{equation}
This background is obtained from the 
${\rm AdS}_5 \times {\rm S}^5$ geometry 
\cite{Blau2,Blau3} via the Penrose limit \cite{Penrose,Guven}. 
 
The action of a scalar field on this background is given by 
\begin{eqnarray}
I_{\rm S} &=& \int\!\! d^{10}x\, \sqrt{-g}\left[ -\frac{1}{2} g^{MN}
      \partial_{M}\phi\,\partial_{N}\phi - V(\phi)\right]  \nn \\
&=& \int\!\! d^{10}x \,
        \left[\:\partial_+\phi\:\partial_-\phi
        - \frac{1}{2} \mu^2 r^2\partial_-\phi\:\partial_-\phi
        - \frac{1}{2}\partial_i\phi\:\partial_i \phi - V(\phi)\right] 
        \,.
        \label{eq;ppaction}
\end{eqnarray}
The effective potential for this system is written as 
\begin{equation}
V_{\rm eff}\left(\phi_{\rm c}\right)
   = V\left(\phi_{\rm c}\right)
    -\frac{i}{2\:V_{10}}{\rm Tr}\ln\left(\Box_{(\rm pp)}
          + V''_{\rm c}\right)\,, 
       \label{eq;ppeffective}
\end{equation}
by using the Laplacian in the pp-wave background 
$\Box_{(\rm pp)}$, which is defined as 
\begin{equation}
\Box_{(\rm pp)} =  2\partial_-\,\partial_+ - \nabla_{(8)}^2 
    - {\mu}^2 \,r^2 \,\partial_-^2\,,
        \label{eq;laplacian}
\end{equation}
where $\nabla_{(8)}^2$ is the one in the transverse eight-dimensional
flat-space.  All we have to do for deriving the effective potential is
to evaluate the trace of the Laplacian. It can be done by the use of the
harmonic-oscillator techniques as discussed in \cite{Mathur}, and then 
the one-loop correction term is given by
\begin{eqnarray}
\hspace{-0.7cm}
-\frac{i}{2V_{10}}\,{\rm Tr}\,\ln \left(\Box_{(\rm pp)} + V''_{\rm c}
          \right)
       &=&-\frac{i}{2V_8}\int^{+\infty}_{-\infty}\!\!\frac{dk_+}{2\pi}
           \int^{\infty}_{-\infty}\!\!\frac{dk_-}{2\pi}
           \sum_{\{n_i\}}\ln
          \left(-2k_+ k_- + |k_-| E_n + V_{\rm c}''\right)\,, 
       \label{eq;ppderi}
\end{eqnarray}
where $E_n$ is defined as 
\begin{equation}
E_n \equiv |\mu| \sum_{i=1}^{8}
                \left(n_i + \frac{1}{2}\right) = |\mu| (n+4)\,, 
\qquad n\equiv \sum_{i=1}^8n_i\,.
        \label{eq;energy}
\end{equation}
We will consider the $\mu>0$ case below. 
Notably, the traces of the discrete spectra do not lead
to the volume factor in contrast with continuous spectra, while the
transverse eight-dimensional space is not compact
(i.e., $V_8$ is infinite). This situation originates from the fact that 
the pp-wave background is not compactified though it has discrete
spectra because of the $(+,+)$-component of the metric. 
Here we will keep the factor $V_8$ and return to this problem later.

We will evaluate the one-loop correction term 
with two methods introduced in the previous section.  

\subsection{Computation with Yan's Formula}

In this subsection we will calculate the effective potential 
with the Yan's formula in
scalar field theories on the pp-wave background. 

Let us evaluate the generalized zeta-function: 
\begin{eqnarray}
\zeta_{\rm Yan}(s)
       &=&\frac{1}{V_8}\sum^{\infty}_{n=0}{}_{n+7}C_7
          \;\int^{\infty}_{-\infty}\frac{dk_+}{2\pi}\left[
           \int^{\infty}_{0}\frac{dk_-}{2\pi}\;
           \left(-2k_+k_- + k_- E_n + V''_{\rm c} \right)^{-s}
           \right.\nn\\
       & & \left.+ \!\!
           \int^{0}_{-\infty}\frac{dk_-}{2\pi}\;
           \left(-2k_+k_- - k_- E_n + V''_{\rm c}\right)^{-s}
           \right]\,,
      \label{eq;pp-yan}
\end{eqnarray}
where the combinational 
factor ${}_{n+7}C_7$ denotes the degeneracy of the sum of 
$n_i$.   We can perform 
the integrals for $k_{+}$ and $k_-$ by using the Yan's formula 
(\ref{eq;yan}). As the result, we obtain the following expression:
\begin{eqnarray}
\zeta_{\rm Yan}(s)
&=& \frac{1}{2(2\pi)^2V_8}
            \sum^{\infty}_{n=0}{}_{n+7}C_7\,\frac{1}{s-1}\,E_n^{-s+1}
            \left\{\int^{\infty}_{0}dk_-
            2\pi i\;\delta(k_-)\;
            \left(k_-+\frac{V_{\rm c}''}{E_n}\right)^{-s+1}
             \right.\nonumber\\
       & & \left.+\int^{\infty}_{0}dk_-
           2\pi i\delta(-k_-)
           \left(k_-+\frac{V_{\rm c}''}{E_n}\right)^{-s+1}
           \right\} \nonumber\\
           &=&i\frac{1}{4\pi V_8}
            \sum^{\infty}_{n=0}{}_{n+7}C_7\,\frac{1}{s-1}
           \left(V_{\rm c}''\right)^{-s+1}\,.
           \label{eq;ppzetayan}
\end{eqnarray}
Using Eqs\,.\,(\ref{eq;sum->zeta}) and (\ref{eq;combi}),
the expression (\ref{eq;ppzetayan}) and its derivative 
can be written as:   
\begin{eqnarray}
\hspace{-0.5cm}\zeta_{\rm Yan}(s)= i\frac{b+1}{4\pi(s-1)V_8}
           \left(V_{\rm c}''\right)^{-s+1}\,,\hspace{0.3cm}
\zeta'_{\rm Yan}(s)=-i\frac{b+1}{4\pi(s-1)V_8}\;
             \left(V_{\rm c}''\right)^{-s+1}
             \left(\frac{1}{s-1}+\ln V_{\rm c}''\right)\,,
\end{eqnarray}
where $b$ is a finite constant defined by
\begin{eqnarray}
b+1=\frac{1}{7!}
  \Big\{\zeta_{\rm R}(-7)-14\zeta_{\rm R}(-5)+49\zeta_{\rm R}(-3)
  -36\zeta_{\rm R}(-1)\Big\}
 =\frac{2497}{3628800}
     \label{eq;a}\,.
\end{eqnarray}
Here we used the Riemann's zeta function $\zeta_{\rm R}(s)$\,.
From the above results, we obtain the formulae:  
\begin{eqnarray}
\lim_{s\rightarrow 0}\zeta_{\rm Yan}(s)
         =-i\:\frac{b+1}{4\pi V_8}\,V_{\rm c}''\,,\hspace{1cm}
\lim_{s\rightarrow 0}\zeta'_{\rm Yan}(s)
        = i\;\frac{b+1}{4\pi V_8}\,V_{\rm c}''
         \left(\ln V_{\rm c}''-1\right)\,.
\end{eqnarray}
Finally the effective potential is given by 
\begin{eqnarray}
V_{\rm eff}\left(\phi_{\rm c}\right)= V\left(\phi_{\rm c}\right)
            -\frac{b+1}{8\pi V_8}\;V_{\rm c}''
           \left\{\ln \left(\frac{V_{\rm c}''}{M^2}\right)-1
           \right\}\,. 
        \label{eq;yanpp-f}
\end{eqnarray}
It should be remarked that the resulting effective potential 
agrees with that in the two-dimensional Minkowski spacetime up to 
the coefficient of the one-loop correction term. Notably, it is also 
independent of the parameter $\mu$\,. 
Mathematically, we can see this fact by noting that 
the effect of $\mu$ (i.e., $E_n$) 
can be absorbed by shifting the light-cone momentum
$k_+$\,. According to this result of the mathematical manipulation,    
the physical interpretation is as follows: 
The vacuum energies produced by the quantum 
fluctuations flow from transverse space to the $k_+$-direction 
due to the flux equipped with the pp-wave geometry. 
In fact, we can see this fact from the classical pp-wave background.  
The non-trivial Einstein equation for the pp-wave geometry 
under our consideration is $R_{++} \sim \mu^2 \sim F^2_{+1234} 
= F^2_{+5678}$, and hence the existence of flux leads to non-vanishing 
curvature only in the $x^+$-direction. 
As the result of the energy flow, the effect of the pp-wave background 
would be realized only as the numerical coefficients, and thus 
the vacuum energy may not explicitly depend on the parameter $\mu$.  

In the next subsection, we will rederive this result by using 
another method.

\subsection{Computation with the Light-Cone Momentum Cut-Off}

Here we will calculate the effective potential by introducing 
the cut-off for the light-cone momentum in the pp-wave case. 

We begin with the evaluation of the one-loop correction term, which 
is rewritten as  
\begin{eqnarray}
{\rm Tr}\,\ln\left\{\Box_{(\rm pp)} + V''_{\rm c}
          \right\}
        =\frac{1}{V_8}
           \int^{+\infty}_{-\infty}\!\!\frac{dk_+}{2\pi}
           \int^{\Lambda}_{V''_{\rm c}/\Lambda}\!\!
           \frac{dk_-}{2\pi}\;\sum_{\{n_i\}}\ln
           \left\{-(2k_+k_-)^2 + (k_- E_n + V_{\rm c}'')^2 \right\}\,.
\end{eqnarray}
Here we define the generalized zeta-function by 
\begin{eqnarray}
\zeta_{\rm pp}(s)=\frac{1}{V_8}\sum^{\infty}_{n=0}{}_{n+7}C_7
           \int^{\Lambda}_{V_{\rm c}''/\Lambda}
          \frac{dk_{-}}{2\pi}\int^{\infty}_{-\infty}\frac{dk_{+}}{2\pi}
          \;\left\{-4k_{+}^2k_{-}^2
        + \left(k_-E_n+V_{\rm c}''\right)^2\right\}^{-s}\,. 
            \label{eq;ppcfzeta}
\end{eqnarray}
Performing the integral for $k_+$ after the
analytic continuation $k_+\rightarrow ik_+$, Eq.\,(\ref{eq;ppcfzeta})
 is rewritten as 
\begin{eqnarray}
\zeta_{\rm pp}(s)&=& i\frac{1}{V_8}
          \sqrt{\pi}\:\frac{\Gamma\left(s-\frac{1}{2}\right)}
           {2(2\pi)^2\,\Gamma(s)}
             \sum^{\infty}_{n=0}{}_{n+7}C_7\,E_n^{-2s+1}
             \int^{\Lambda}_{V_{\rm c}''/\Lambda}\frac{dk_{-}}{k_{-}}
             \left(k_-+\frac{V_{\rm c}''}{E_n}\right)^{-2s+1}\,.
          \label{eq;zetacfpp1}
\end{eqnarray}
Then the generalized zeta-function is explicitly written as  
\begin{eqnarray}
\zeta_{\rm pp}(s)
      &=& \frac{i}{V_8}
          \sqrt{\pi}\:\frac{\Gamma\left(s-\frac{1}{2}\right)}
             {2(2\pi)^2\Gamma(s)}
             \sum^{\infty}_{n=0}{}_{n+7}C_7\,E_n^{-2s+1}\,
             \frac{1}{2s-1}\nn\\
      & & \quad 
          \times\left\{
             {}_2F_1\left(2s-1,\;2s-1,\;2s,\;-\frac{\Lambda}{E_n}\right)
              \left(\frac{V_{\rm c}''}{\Lambda}\right)^{-2s+1}
               \right.\nonumber\\
      & & \quad \left.
            -{}_2F_1\left(2s-1,\;2s-1,\;2s,\;
             -\frac{V_{\rm c}''}
             {\Lambda\;E_n}\right)\;\Lambda^{-2s+1}\right\}\,.
      \label{eq;calzeta2}
\end{eqnarray}
The hypergeometric function ${}_2F_1$ 
can be expanded\footnote{We comment on the convergence radius
of the expansion with respect to $1/\mu$\,.  In the large $\mu$\,, 
the hypergeometric function ${}_2F_1$
is expanded in terms of the $\Lambda/\mu$ and $V''_{\rm c}/(\Lambda\mu)$\,. 
Then $\zeta_{\rm pp}$ converges for the case 
$\Lambda<\mu$ and $V''_{\rm c}<\Lambda\mu$\,. } 
as 
\begin{eqnarray}
{}_2F_1\left(2s-1,\;2s-1,\;2s,\;
             -\frac{\Lambda}{E_n}\right)
            &=&1-\frac{(2s-1)^2}{2s}\frac{\Lambda}{E_n}
             +\frac{s(2s-1)^2}{2s+1}\left(\frac{\Lambda}{E_n}\right)^2
                 \nonumber\\ 
          & &\hspace{-2cm} -\frac{s(2s-1)^2(2s+1)}{3(2s+2)}
              \left(\frac{\Lambda}{E_n}\right)^3
             +O\left\{\left(\frac{\Lambda}{\mu}\right)^4\right\}\,,
         \\
{}_2F_1\left(2s-1,\;2s-1,\;2s,\;
             -\frac{V_{\rm c}''}{\Lambda\;E_n}\right)
            &=&1-\frac{(2s-1)^2}{2s}\frac{V_{\rm c}''}{\Lambda\;E_n}
          +\frac{s(2s-1)^2}{2s+1}\left(\frac{V_{\rm c}''}
           {\Lambda\;E_n}\right)^2
                 \nonumber\\ 
          & &\hspace{-2cm} -\frac{s(2s-1)^2(2s+1)}{3(2s+2)}
              \left(\frac{V_{\rm c}''}{\Lambda\;E_n}\right)^3
             +O\left\{\left(\frac{V_{\rm c}''}
             {\Lambda\;\mu}\right)^4\right\}\,, 
        \label{eq;expand}
\end{eqnarray}
and the generalized zeta-function
 (\ref{eq;calzeta2}) can be rewritten as 
\begin{eqnarray}
&& \hspace{-1.2cm}\zeta_{\rm pp}(s) \;=\; \frac{i\sqrt{\pi}}{V_8}
       \left[\frac{1}{2(2\pi)^2}
                \frac{\Gamma\left(s-\frac{1}{2}\right)}
       {(2s-1)\;\Gamma(s)}\sum^{\infty}_{n=0}{}_{n+7}C_7\:
             \left\{\mu(n+4)\right\}^{-2s+1}
             \left\{\left(\frac{V_{\rm c}''}{\Lambda}\right)^{-2s+1}
              -\Lambda^{-2s+1}\right\}
              \right.
                 \nonumber\\  
      & &-\frac{1}{2(2\pi)^2}\frac{2s-1}{2s}
                \frac{\Gamma\left(s-\frac{1}{2}\right)}{\Gamma(s)}
                \sum^{\infty}_{n=0}{}_{n+7}C_7\:
                \left\{\mu(n+4)\right\}^{-2s}V_{\rm c}''
                \left\{\left(\frac{V_{\rm c}''}{\Lambda}\right)^{-2s}
                 -\Lambda^{-2s}\right\}
                 \nonumber\\  
      & & \left. +O\left(\frac{1}{\mu^{2s+1}}\right)\right]\,.
        \label{eq;rezeta2}
\end{eqnarray}
Using Eqs.\,(\ref{eq;rezeta2}), (\ref{eq;sum->zeta})  and 
(\ref{eq;sum->zeta1})  leads to the limits: 
\begin{eqnarray} 
\lim_{s\rightarrow 0}\zeta_{\rm pp}(s)
      =0\,, \hspace{1cm} 
\lim_{s\rightarrow 0}\zeta'_{\rm pp}(s)
         = i\;\frac{b+1}{4\pi V_8}
         \;V_{\rm c}''\ln\left(\frac{V_{\rm c}''}{\Lambda^2}\right)
          \,.
        \label{eq;zetaderi2}   
\end{eqnarray}
Here $b$ is defined in Eq.\,(\ref{eq;a}).
It should be noted that higher order terms $O(1/\mu)$
 vanish in the limit $s \rightarrow 0$. 

Finally, the effective potential is represented by 
\begin{eqnarray}
V_{\rm eff}\left(\phi_{\rm c}\right)
           =V\left(\phi_{\rm c}\right)
           -\frac{b+1}{8\pi V_8} \,
           V_{\rm c}''\ln \left(\frac{V_{\rm c}''}{\Lambda^2}
           \right)\,.
            \label{eq;zetappl}
\end{eqnarray}
When we define the scale parameter $M$ in
Eq.\,(\ref{eq;yanpp-f}) as $M^2=\Lambda^2\,{\rm e}$\,,
 the above expression becomes
identical with the previous result derived by using the Yan's formula.

Now let us comment on the vacuum energy. We consider the quadratic
potential of scalar field defined by $V(\phi)=m^2\phi^2/2$\,. 
By inserting it into the effective potential (\ref{eq;yanpp-f}) 
(or (\ref{eq;zetappl})), we obtain 
\begin{eqnarray}
V_{\rm eff}\left(\phi_{\rm c}\right)
           =\frac{1}{2}m^2\phi_{\rm c}^2
            -\frac{b+1}{8\pi V_8}\;m^2
            \left\{2\ln \left(\frac{m}{M}\right)-1\right\}\,,
            \label{eq;effm^2}
\end{eqnarray}
where the second term expresses the one-loop correction and may be
interpreted as the vacuum energy. 
It is well-known that the zero-point energy of massive fields 
may contribute to the cosmological constant. 

\begin{figure} 
\begin{center}
\includegraphics[width=10cm,clip]{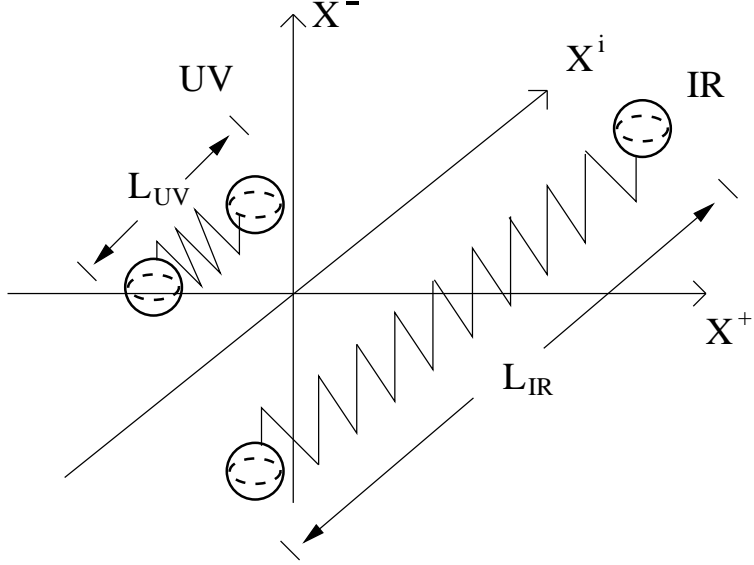} \caption{\footnotesize
This figure shows the ``effective'' compactification for the transverse
directions.  The effective scale for the transverse direction is
evaluated by $(\mu\,k_-)^{-1/2}$.  Then the moving region at the scale
$\Lambda$ (UV) is far shorter than that of $V''_{\rm c}/\Lambda $ (IR)
when the cut-off scale $\Lambda$ is assumed to be large enough.}
\label{fig;1}
\end{center}
\end{figure}

From now on, we discuss the volume of eight-dimensional transverse
space. It would be plausible to measure the volume with the damping
length of the wave packet in the system of harmonic oscillators.  That
is, the volume $V_8$ is regarded as the moving region of harmonic
oscillator (See Fig.\,\ref{fig;1}).
 This length is estimated as $(\mu\:k_-)^{-1/2}$\,.  When we assume that
the cut-off scale for $k_-$ is imposed by $V''_{\rm c}/\Lambda \le |k_-|
\le \Lambda$\,, the effective potential for the infrared cut-off
$V''_{\rm c}/\Lambda$ is
\begin{eqnarray}
V_{\rm eff}\left(\phi_{\rm c}\right)&\sim& V\left(\phi_{\rm c}\right)
            -\frac{b+1}{8\pi}\;\left(\frac{\mu}{\Lambda}\right)^4
           \left(V_{\rm c}''\right)^5
           \left\{\ln \left(\frac{V_{\rm c}''}{\Lambda^2}\right)-1
           \right\}\,. 
\end{eqnarray}
This is just an effective potential of scalar field in ten-dimensional
Minkowski spacetime.  On the other hand, the effective potential for
ultraviolet cut-off $\Lambda$ is given by
\begin{eqnarray}
V_{\rm eff}\left(\phi_{\rm c}\right)&\sim& V\left(\phi_{\rm c}\right)
            -\frac{b+1}{8\pi}\:\left(\mu\,\Lambda\right)^4\;V_{\rm c}''
           \left\{\ln \left(\frac{V_{\rm c}''}{\Lambda^2}\right)-1
           \right\}\,. 
\end{eqnarray}
This result is equivalent to that in two-dimensional Minkowski
spacetime.

Finally we shall briefly comment on the limit $\mu \rightarrow 0$\,. 
We can recover the effective potential in ten-dimensional 
Minkowski spacetime from (\ref{eq;ppderi})
by following the replacement law: 
\begin{eqnarray}
\mu\:k_-\left(n_i+\frac{1}{2}\right) \Rightarrow k_i^2\,, \qquad 
\frac{1}{V_8}
 \sum_{\{n_i\}} \Rightarrow \int \frac{d^8 k}{(2\pi)^8}\,.
\end{eqnarray} 
What does it physically mean? 
If we take the limit $\mu \rightarrow 0$ naively, then we 
cannot recover the transverse momentum. We need to take account 
of the energy conservation of the harmonic oscillations. 
The above relations mean that the energy of the harmonic oscillators
should be replaced with the kinematic energy of free motion.   
In other words, we have to take account of the Casimir energy.

Once we suppose the above replacement law, 
we can easily obtain
the effective potential in ten-dimensional Minkowski spacetime 
\begin{eqnarray}
V_{\rm eff}\left(\phi_{\rm c}\right)
         = V\left(\phi_{\rm c}\right)
          -\frac{1}{15\cdot2^4(4\pi)^{5}}
          \left(V''_{\rm c}\right)^5
          \left\{\ln\left(\frac{V''_{\rm c}}{\Lambda^2}
          \right)-\frac{137}{60}\right\}\,,
\end{eqnarray}
by using 
\begin{eqnarray}
\lim_{s\rightarrow 0}
  \left\{\frac{\Gamma\left(s-5\right)}{\Gamma(s)}\right\}
   =-\frac{1}{120}\,,
  \quad \frac{\Gamma\left(s-5\right)\Big\{\psi(s-5)
   - \psi(s)\Big\}}{\Gamma(s)}
   =\frac{1}{120}\left(-\frac{137}{60}\right)+O(s)\,. 
\end{eqnarray}

\section{Conclusion and Discussion} 

We have developed two methods to calculate 
effective potentials in the light-cone frame 
for the application in the pp-wave case. 
One is to use the Yan's formula and the other is to introduce 
a cut-off for the light-cone momentum. 
These methods would be available for the study of the usual 
theories such as QED and QCD in the light-front formulation 
other than quantum field theories on pp-wave backgrounds.  

By using our two methods, we have calculated the effective potential in
a scalar field theory on the ten-dimensional maximally supersymmetric
pp-wave background.  The effective potential obtained here is similar to
that in two-dimensional Minkowski spacetime up to the coefficient of the
one-loop correction term. It would be an important subject to compute
effective potentials in the canonical formulation of the discrete light-cone
quantization (DLCQ)\footnote{As an example of the works 
for such a direction in flat space, we can refer to
\cite{Heinzl:2000ht}.}  
and to confirm our result in this paper, while we have utilized the path
integral formulation in this paper. 
We will study in this direction and report in the near future \cite{UY2}.

There are many future problems. 
The pp-wave background considered here is static and the 
scalar field theory on this background 
is not realistic as a cosmological model. Thus, 
it is one of the most interesting problems to 
consider scalar field theories on time-dependent 
plane-wave backgrounds as in
\cite{Papadopoulos:2002bg,BO,BOPT,BMO,SaYo}. 
Also, it might be possible to apply our methods  
to the case of time-dependent orbifolds. Moreover,  
it is nice to consider scalar field theories on 
 {\it four-dimensional} pp-wave backgrounds. 

In addition, it is interesting to study effective potentials in scalar
field theories on plane-wave backgrounds obtained from (AdS) black holes
via Penrose limits \cite{Hubeny,Brecher:2002bw,Marolf:2002bx,Li}.  In
these cases we would face a harmonic oscillator with a negative
mass-square, and hence the behavior of the sum of modes (i.e., one-loop
corrections) would be drastically modified.  It has been, however,
argued in the literatures \cite{Hubeny,Brecher:2002bw, Marolf:2002bx}
that such a negative mass square would not lead to an imaginary part in
the effective potential and hence a break-down of unitarity would not be
caused.  It would be an attractive work to check this argument from the
viewpoint of one-loop effective potential in order to promote our
understanding of pp-wave geometry and Penrose limit.

\section*{Acknowledgment}

 We would like to thank M. Sakagami and K. Sugiyama for 
continuing encouragement and helpful discussion. 
We also thank T. Uematsu for useful comments.

\vspace*{1cm}

\newpage
\noindent
{\large{\bf Appendix}}

\appendix

\section{Useful Formulae}
  \label{sec;appenA}

In this appendix A, we shall summarize useful formulae 
for the computation of effective potential 
by using the zeta function regularization method. 
These are surely available for the calculation in section \ref{sec;pp}. 

\subsection*{Integration Formula}

In order to compute effective potentials, 
we first use the following integration formulae: 
\begin{eqnarray}
& &\hspace{-2cm}
   \int^{\infty}_{-\infty}\!d^dx\,\left(x^2+M^2\right)^{-s}
     = \pi^{d/2}\,
       \frac{\Gamma\left(s-\frac{d}{2}\right)}{\Gamma(s)}
       \left(M^2\right)^{-s+d/2}\,,\nn\\
    \label{eq;int1}
& &\hspace{-2cm}
     \int^{\Lambda_2}_{\Lambda_1}\!dx\,\frac{\left(x+M^2\right)^{-s}}{x}
         =\frac{1}{s}\Bigl\{\Lambda_1^{-s}\,
            {}_2F_1\left(s,\;s,\;s+1,\;-\frac{M^2}{\Lambda_1}\right)
         \nn\\
& &
       -\Lambda_2^{-s}\,
            {}_2F_1\left(s,\;s,\;s+1,\;-\frac{M^2}{\Lambda_2}\right)
       \Bigr\}\,,
    \label{eq;int2}           
\end{eqnarray}
where ${}_2F_1\left(\alpha,\;\beta,\;\gamma,\;z\right)$ is the Gauss's 
hypergeometric function:
\begin{eqnarray}
{}_2F_1(\alpha,\;\beta,\;\gamma,\;z)
     &=&\frac{\Gamma(\gamma)}{\Gamma(\alpha)\:\Gamma(\beta)}
      \sum^{\infty}_{n=0}
      \frac{\Gamma(\alpha+n)\:\Gamma(\beta+n)}{\Gamma(\gamma+n)}
      \frac{z^n}{n!}\,.
       \label{eq;hypergeometric}    
\end{eqnarray}

\subsection*{Formulae of Summation}

In the calculation of effective potential in 
scalar field theories on a pp-wave background, 
we have to evaluate the following infinite sum:
\begin{eqnarray}
\sum^{\infty}_{n=0}{}_{n+7}C_7 \left(n+4\right)^{-s-d}\,.
\end{eqnarray}
The sum without the combinatorial factor can be rewritten as   
\begin{eqnarray}
\hspace{-0.5cm}\sum^{\infty}_{n=0}\left(n+4\right)^{-s-d}
   &=&\zeta_{\rm R}(s+d)-1-2^{-s-d}-3^{-s-d}+4^{-s-d}\,,
        \label{eq;sum->zeta} 
\end{eqnarray}
by using the Riemann's zeta function $\zeta_{\rm R}(s)$ defined as 
\begin{eqnarray}
\zeta_{\rm R}(s)&=&\sum^{\infty}_{n=1}n^{-s}\,.
\end{eqnarray}
On the other hand, noting that 
the combinatorial factor ${}_{n+7}C_7$ can be represented by  
\begin{eqnarray}
{}_{n+7}C_7 
&=& 
\frac{1}{7!}\left\{\left(n+4\right)^7-14\left(n+4\right)^5
       +49\left(n+4\right)^3-36\left(n+4\right)\right\}\,,
   \label{eq;combi}  
\end{eqnarray}
one can obtain the following relation:
\begin{eqnarray}
\sum^{\infty}_{n=0}{}_{n+7}C_7\:(n+4)^{-s-d} 
&=& \frac{1}{7!}\left\{
                \zeta_{\rm R}\Big(s+(d-7)\Big)
               -14\zeta_{\rm R}\Big(s+(d-5)\Big)
\right. \nn \\
&& \left. +49\zeta_{\rm R}\Big(s+(d-3)\Big)
   -36\zeta_{\rm R}\Big(s+(d-1)\Big)
   +4^{-s-d}\right\}\,.
        \label{eq;sum->zeta1}
\end{eqnarray}
This formula is also useful in the calculation in section \ref{sec;pp}.

\end{document}